\documentclass[final]{elsart}
\newcommand{\avrg}[1] {\left<  #1  \right>}

\usepackage{color}

\usepackage{epsfig}
\usepackage{slashbox}
\usepackage{lineno}
\usepackage{amssymb}
\usepackage{lineno}

\begin{document}

\begin{frontmatter}



%\title{Determination of an asymmetry parameter}
\title{Comparison of methods to extract an asymmetry parameter from data}


\author{J\"org Pretz\corauthref{cor1}}
 \corauth[cor1]{corresponding author}
\ead{jorg.pretz@cern.ch}
\address{Physikalisches Institut, Universit\"at Bonn, 53115 Bonn, Germany}

\begin{abstract}
Several methods to extract an asymmetry parameter in an event distribution function are 
discussed and compared 
in terms of statistical precision and applicability.
These methods are: simple counting rate asymmetries,
event weighting procedures and 
the unbinned extended maximum likelihood method. 
It is known that weighting methods reach the same
figure of merit (FOM) as the likelihood method in the limit of vanishing asymmetries.
This article presents an improved weighting procedure reaching the FOM 
of the likelihood method for arbitrary asymmetries.
Cases where the maximum likelihood method is not applicable are also discussed.
\end{abstract}
%in the limit of vanishing asymmetries.
%Reference \cite{Pretz:2008mi} showed that using event 
%weighting, one can reach the same figure of merit (FOM) as the likelihood method 
%in the limit of vanishing asymmetries.

\begin{keyword}
event weighting \sep minimal variance bound \sep Cram\'er-Rao inequality \sep asymmetry  extraction
\sep optimal observables \sep parameter determination  \sep maximum likelihood 

\PACS 02.70.Rr \sep 13.88.+e 
\end{keyword}
\end{frontmatter}

% main text

%-------------------------------------------------------------------------------
\section{Introduction}
We consider two differential event distributions $n^\pm(x)$ following 
the functional form
\begin{equation}\label{N1}
  n^\pm(x) = \alpha(x) (1 \pm \beta(x) A)  \, .
\end{equation}
In a typical experimental situation encountered in particle physics $\alpha(x)$ includes a flux and acceptance factor and $\beta(x)$ is an analyzing power. 
Both depend on a set of kinematic variables here denoted by $x$.
Concrete examples are spin cross section asymmetries and muon decay.
In the latter case the asymmetry corresponds to the muon polarization.
The two data sets ($+$ and $-$) are for example obtained by changing the sign 
of a polarization.
The goal is to extract the parameter $A$ by measuring the event distributions
$n^\pm(x)$. This would be an easy task if both  $\alpha(x)$ and $\beta(x)$ were known.
However in many applications the factor $\alpha(x)$ is not known, or not
accurately enough known and only $\beta(x)$ is given.

Section~\ref{diff_methods} presents various methods to extract the parameter $A$:
The simplest method, based on counting rate asymmetries, 
the more efficient extended unbinned maximum likelihood (EML) method and finally
methods based on event weighting are discussed.
While in Section~\ref{diff_methods} we assume that the factor $\alpha(x)$ is the same for both data sets,
Section~\ref{cne1} extends the discussion to the case where one has two
different factors.
%the flux and acceptance factor $\alpha$ is not the same in the  two data sets.
Up to Section~\ref{cne1} we assume that the number of observed events is large enough
so that averages can be replaced by the corresponding expectation values. 
Effects occurring at low statistics event samples are discussed in
Section~\ref{low_stat}.
A summary and conclusions are given in Section~\ref{sum}.

\section{Different Methods to extract $A$}\label{diff_methods}
\subsection{Determining $A$ from counting rate asymmetry}\label{a_cnt}
The expectation value of the number of events for the two data sets reads
\begin{equation}\label{Nplmi}
  \avrg{N^{\pm}} = \int n^\pm(x) {\rm d} x = \left( 1 \pm \avrg{\beta} A \right) \int \alpha(x) {\rm d} x 
\end{equation}
with $\avrg{\beta} = {\int \alpha \beta {\rm d} x}/{\int \alpha {\rm d} x}$.
The integrals run over the kinematic range of $x$.
The asymmetry $A$ can be extracted without the knowledge of $\int \alpha(x) {\rm d} x$:
\begin{equation}\label{A_cnt}
 A = \frac{1}{\avrg{\beta}} \,
 \frac{\avrg{N^{+}}-\avrg{N^{-}}}{\avrg{N^{+}}+\avrg{N^{-}}} \, .
\end{equation}
Eq.~(\ref{A_cnt}) leads to following estimator 
for $A$:
\[ 
 \hat A_{cnt} =\frac{N^+ + N^-}{\sum_{+}\beta_i + \sum_{-} \beta_i}\, \,
 \frac{N^+-N^-}{N^++N^-} = 
\frac{N^+-N^-}{\sum_+ \beta_i + \sum_- \beta_i}
\]
where $\beta_i \equiv \beta(x_i)$, $N^+$ and $N^-$ are the numbers of observed events.
The sums $\sum_{+}$ and $\sum_{-}$ run over all events in the corresponding data set ($+$ or $-$).

As shown in Section~\ref{wm} and App.~\ref{app_covwb}, the figure of merit
(FOM), i.e. the inverse of the variance on $\hat A_{cnt}$, is
\begin{equation}\label{a1}
\mbox{FOM}_{\hat A_{cnt}} = N \frac{\avrg{\beta}^2}{1-A^2 \avrg{\beta^2}} 
\end{equation}
where $N = N^+ + N^-$ denotes the total number of events.

This figure of merit may be increased if a cut is set to remove some data
with low values of $\beta$.  However, it will not reach the FOM of the
unbinned extended 
likelihood method discussed now, unless $\beta(x)$ is constant.

\subsection{Extended Maximum Likelihood (EML) Method}
We now turn to the unbinned extended maximum likelihood (EML) method\cite{barlow,barlow_lh} which is known to reach the Cram\'er-Rao limit of the lowest possible statistical error.
The log-likelihood function derived from Eq.~(\ref{N1}) reads
\begin{eqnarray}
  l &=& \sum_+ \ln \left(\alpha_i (1 + \beta_i A) \right)- \langle N^+\rangle (A)  \nonumber 
                  + \sum_- \ln \left(\alpha_i (1 - \beta_i A) \right)- \langle N^-\rangle (A)  \, .\nonumber
\end{eqnarray}
Using the expression in Eq.~(\ref{Nplmi}) for the expectation values
$\avrg{N^{\pm}}$ results in
\begin{eqnarray}
  l &=& \sum_+ \ln \left(1 + \beta_i A \right)  + \sum_- \ln \left(1 -
  \beta_i A \right)  - 2 \int \alpha(x) {\rm d} x - \sum_{+,-} \ln \alpha_i \, .
\end{eqnarray}
The last two terms do not
depend on $A$ and can be ignored
in the likelihood maximization. The asymmetry $A$ can thus be determined without knowledge of $\alpha(x)$.
For small values of $\beta A$ one can even derive an analytic expression for $\hat A_{LH}$
which reads
\begin{equation}\label{alh}
 \hat A_{LH} = \frac{\sum_+ \beta_i - \sum_- \beta_i}{\sum_+ \beta^2_i +
   \sum_- \beta^2_i} \, .
\end{equation}
For arbitrary asymmetries the maximization has to be done numerically.
Note, that this requires CPU intensive sums over all events in the
maximization procedure.

The figure of merit (FOM) is
given by
\[
\mbox{FOM}_{\hat A_{LH}} =- \frac{\partial^2 l}{\partial A^2} = \sum_+
\frac{\beta_i^2}{(1+\beta_i A)^2} + \sum_- \frac{\beta_i^2}{(1-\beta_i A)^2}
\, .
\] 
Replacing the sum over events by integrals one finds
\begin{eqnarray}
  \mbox{FOM}_{\hat A_{LH}} &=& \int \frac{\alpha (1+\beta A) \beta^2}{(1+\beta A)^2} {\rm d} x +
  \int \frac{\alpha (1-\beta A) \beta^2}{(1-\beta A)^2} {\rm d} x \nonumber \\
         %        &=& \int \alpha \beta^2 \left( \frac{1}{1+\beta A} + \frac{1}{1-\beta A}\right) \\
           &=& \int \frac{2 \alpha \beta^2}{1-\beta^2 A^2} {\rm d} x\, .
\end{eqnarray}
Noting, that for an arbitrary function $f(x)$ the average is defined by
\[
  \avrg{f} = \frac{\int f(x) (n^+(x) + n^-(x)) {\rm d} x}{\int n^+(x) + n^-(x) {\rm d}
    x} = \frac{\int \alpha(x) f(x){\rm d} x }{\int \alpha(x) {\rm d} x}\, ,
\]
the figure of merit can be written as 
\begin{equation}\label{fom_lh}
  \mbox{FOM}_{\hat A_{LH}} = N \left<\frac{\beta^2}{1-\beta^2 A^2}\right> \, .
\end{equation}

\subsection{Weighting Method}\label{wm}
Next, consider the following estimator 
\begin{equation}\label{aw}
\hat A_w = \frac{\sum_+ w_i- \sum_- w_i}{\sum_+ w_i \beta_i  + \sum_- w_i \beta_i}
\end{equation}
where $w_i \equiv w(x_i)$ is a, for the moment arbitrary, weight factor assigned to
every event.
The expectation value of $\hat A_w$ equals $A$ independently of the weight function $w(x)$ used.
App.~\ref{app_covwb} shows that 
\begin{equation}\label{fom_aw}
  \mbox{FOM}_{\hat A_w} %= \frac{(\sum w_i \beta_i)^2}{\sum w_i^2 - A^2 \sum w^2 \beta^2} 
%= N \frac{\avrg{w\beta}^2}{\avrg{w^2} - A^2 \avrg{w^2 \beta^2}} \, 
= N \frac{\avrg{w\beta}^2}{\avrg{w^2 (1-A^2 \beta^2)}}.
\end{equation}

Two cases are of interest: \\
1.) Setting $w=1$  corresponds to the counting rate asymmetry discussed in
Section~\ref{a_cnt} and proves Eq.~(\ref{a1}) for the FOM.

2.) In the case $w=\beta$ the FOM is
\[
\mbox{FOM}_{\hat A_{w=\beta}} = N \frac{\avrg{\beta^2}}{1 - A^2 \frac{\avrg{\beta^4}}{\avrg{\beta^2}} }.
\]
A comparison with Eq.~(\ref{fom_lh}) indicates that $\mbox{FOM}_{\hat A_{w=\beta}}$
coincides with the FOM of the likelihood method for vanishing $A$.
Actually, in this case, the two estimators are identical as can be seen by comparing Eq.~(\ref{aw}) with $w \equiv \beta$ and Eq.~(\ref{alh}).
Note that the estimator $\hat A_{w}$ can be applied for arbitrary asymmetries
as well, accepting a decrease of the FOM compared the EML estimator as
discussed in Section~\ref{comp}.

Such a weighting procedure has been used for example in
Refs.~\cite{smc,compass} to extract spin asymmetries in the case where
$\avrg{\beta} A \ll 1$.
In Ref.~\cite{Pretz:2008mi} a weighting method is discussed to 
simultaneously extract signal and background asymmetries.
The fact that a weighting procedure reaches the same FOM as the EML method was first
discussed in Ref.~\cite{barlow1} in the context of signal and background
extractions.
The next section shows that one can find a weight factor reaching the FOM of the EML method
even in the case of non-vanishing asymmetries.

\subsection{Improved Weighting Method}
Variational calculus shows (s. App.~\ref{vc}) that the maximum FOM is reached
using a weight factor
\begin{equation}\label{iw}
 w = \frac{\beta}{1-\beta^2 A_0^2} \, .
\end{equation}
Here, $A_0$ is a first estimate of the asymmetry $A$ obtained
for example from the weighting method presented in Section~\ref{wm}.
The weighting factor defined in Eq.~(\ref{iw}) leads to the following estimator (the index $iw$ stands for improved weight)
\begin{eqnarray}
\hat A_{iw} &=& \frac{\sum_+ \frac{\beta_i}{1-\beta_i^2 A_0^2}- \sum_- \frac{\beta_i}{1-\beta_i^2 A_0^2}}
      {\sum_+ \frac{\beta_i^2}{1-\beta_i^2 A_0^2}+ \sum_- \frac{\beta_i^2}{1-\beta_i^2 A_0^2}} \, . \label{a_iw}
\end{eqnarray}

Eq.~(\ref{fom_aw}) gives
\begin{eqnarray}
  \mbox{FOM}_{\hat A_{iw}} &=& N \, \frac{\avrg{\frac{\beta^2}{1-\beta^2 A_0^2} }^2}{\avrg{\beta^2
      \frac{1-\beta^2 A^2}{(1-\beta^2 A_0^2)^2}}} \, . \label{fom_iw}
\end{eqnarray}
Thus given a good estimate $A_0 \approx A$,
we get $\mbox{FOM}_{\hat A_{iw}}$ = $\mbox{FOM}_{\hat A_{LH}}$, i.e.
the improved weighting method reaches the same FOM as the EML method
for arbitrary asymmetries as well with the advantage 
that no CPU consuming maximization procedure is needed.

Before we move to a comparison of the different methods, we note that 
the estimator
\begin{eqnarray}
 \hat A &=&  \frac{\sum_+ \beta^+_i - \sum_- \beta^-_i}{\sum_+ (\beta^+_i)^2 +
    \sum_- (\beta^-_i)^2} + A_0 \quad \mbox{with} \nonumber \\  
\beta^\pm &=&   \frac{\beta}{1 \pm \beta A_0}  \label{ow1}
\end{eqnarray}
reaches as well the FOM of the EML for $A \approx A_0$,
In contrast to $\hat A_{iw}$ its expectation value only equals $A$ if $A_0\approx A$.
For $\tau$ decays the optimal weight factor in Eq.~(\ref{ow1}) is discussed in Ref.~\cite{davier}.

\subsection{Comparison of different methods}\label{comp}
\begin{figure}
\includegraphics[width=\textwidth]{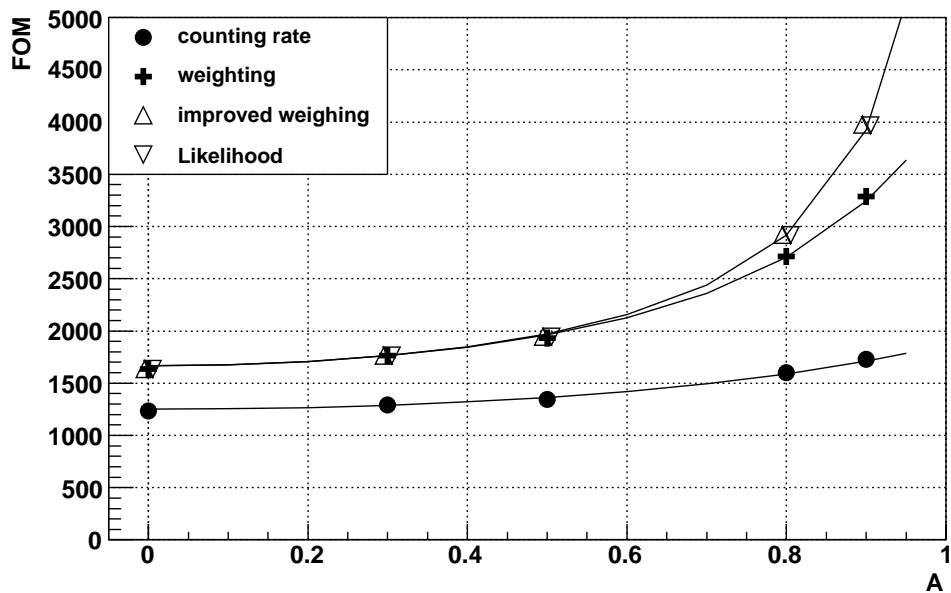}
\caption{Figure of merit for different methods as a function of the
  asymmetry $A$. \label{fig_fom}}
\end{figure}

Tab.~\ref{tab_fom} summarizes the FOM of the various estimators proposed.
\begin{table}
\begin{tabular}{|l|l|l|l|l|}
\hline
            &Counting rate &  weighting & Likelihood & Improved weighting\\
            &asymmetry  &  method    & method     &  method\\
\hline
            &           &               &   \multicolumn{2}{|c|}{}                       \\
$\frac{\mbox{Figure of merit}}{N}$ & $\frac{ \avrg{\beta}^2}{1-A^2 \avrg{\beta^2}}$  & 
$\frac{\avrg{\beta^2}}{1-A^2 \frac{\avrg{\beta^4}}{\avrg{\beta^2}}}$   
  &    \multicolumn{2}{|c|}{ $\avrg{\frac{\beta^2}{1-\beta^2 A^2}}$}    \\
            &           &               &       \multicolumn{2}{|c|}{}                    \\
\hline
\end{tabular}
\caption{The ratio FOM/$N$ for various methods discussed.\label{tab_fom}}
\end{table}
Fig.~\ref{fig_fom} shows the figure of merit of the
different estimators vs. $A$ for the choice
\begin{eqnarray}
 \alpha(x) = \mbox{const.} = 2500  \quad & \mbox{and} & \quad \beta(x) = x
 ,\quad   0<x<1 \, . \label{alpha}
\end{eqnarray}
The curves are analytic calculations.
The points are results of simulations. 
For each value of the asymmetry 10000 configurations with $\alpha=2500$,
which corresponds on average to 5000 events, were simulated.
One configuration consists of a plus and minus data set used to evaluate an asymmetry.
For each of the 10000 configurations simulated, the asymmetries were calculated using
the estimators discussed above. The FOM was determined from the RMS
of the asymmetry distributions. 

The results are in perfect agreement with the analytic calculations.
The statistical errors of the simulations are of the order of the size of the points.
Note, that for all methods no bias was found for the asymmetry.
The question of bias and the range of validity of the expressions for the
FOM will be discussed in more detail in Section~\ref{low_stat}.
The weighting methods are superior to the method using simply the counting
rates. As expected the improved weighting or the EML method reach a higher
FOM than the simple weighting method the larger the asymmetry $A$. 
The results depend of course on the shape of $\alpha(x)$ and $\beta(x)$.
The gain in FOM using weighted events compared to counting rates depends on the spread of $\beta(x)$.
For $A=0$ for example it is $\avrg{\beta^2}/\avrg{\beta}^2$ as can be derived
from Eq.~(\ref{fom_aw}). 

\begin{figure}
\includegraphics[width=\textwidth]{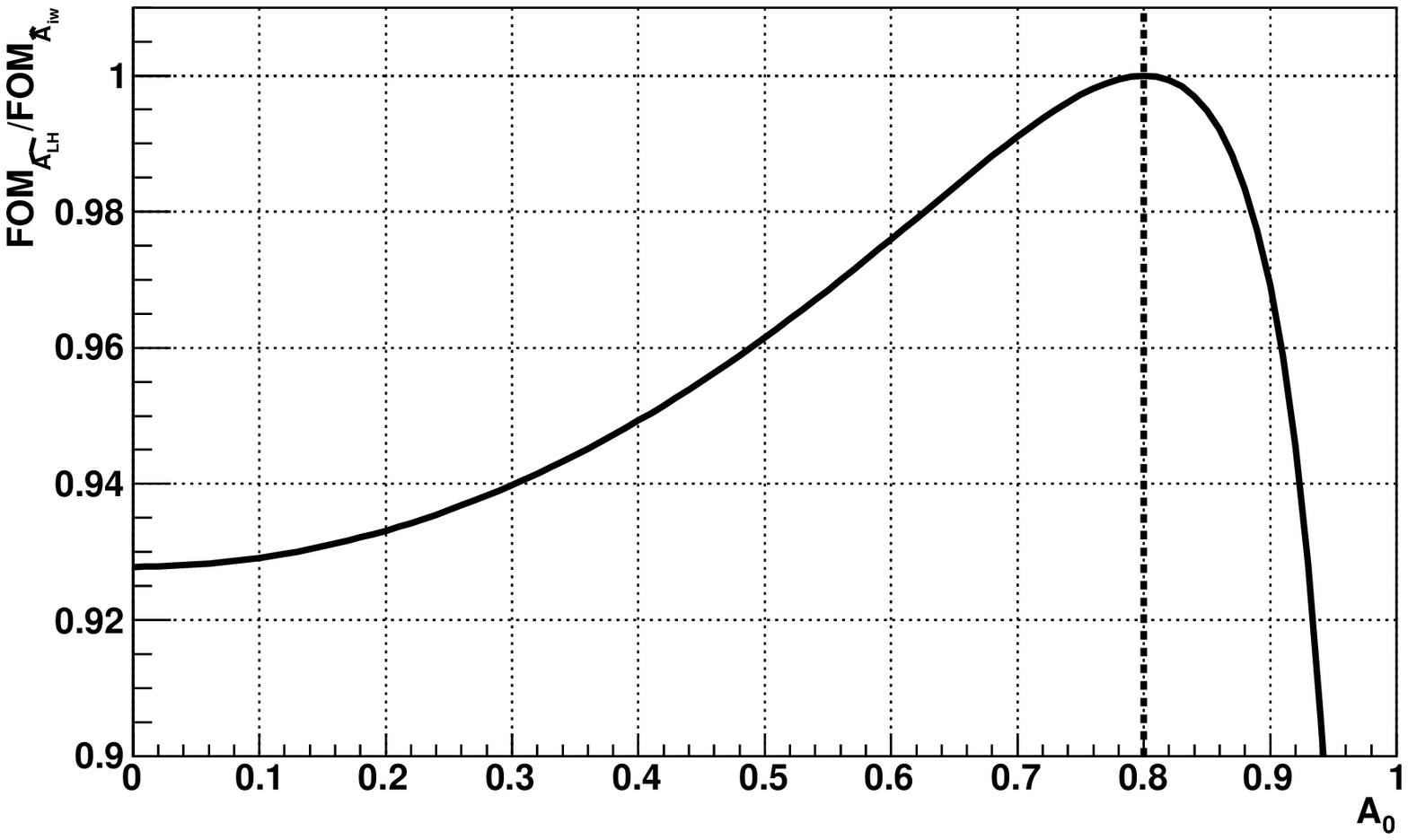}
\caption{$\mbox{FOM}_{\hat A_{iw}}/\mbox{FOM}_{\hat A_{LH}}$ for $A=0.8$ as a function of $A_0$. \label{fig_fom_iw}}
\end{figure}

Fig.~\ref{fig_fom_iw} shows the influence of the choice of $A_0$ on the FOM
in the improved weighting method for the factors
$\alpha$ and $\beta$ as given in Eq.~(\ref{alpha}) and an asymmetry $A=0.8$.
Choosing $A_0$ in a range 0.7--0.86, one reaches at least 99\% of FOM$_{\hat A_{LH}}$.
The normal weighting method corresponds to $A_0=0$.

\section{Different acceptance/flux factor in the two data sets}\label{cne1}
We now turn to the case where
the acceptance and flux factor $\alpha$ is not the same 
in the two data sets. We assume that they differ by a known factor $c$ 
which is independent of $x$.
In this case the differential event distributions are given by
\begin{eqnarray}
 n^+(x) &=& \frac{2c}{1+c} \, \alpha(x) (1+\beta(x) A) \, \quad \mbox{and}\\
 n^-(x) &=& \frac{2}{1+c} \, \alpha(x) (1-\beta(x) A) \, . 
\end{eqnarray}

\begin{figure}
\includegraphics[width=\textwidth]{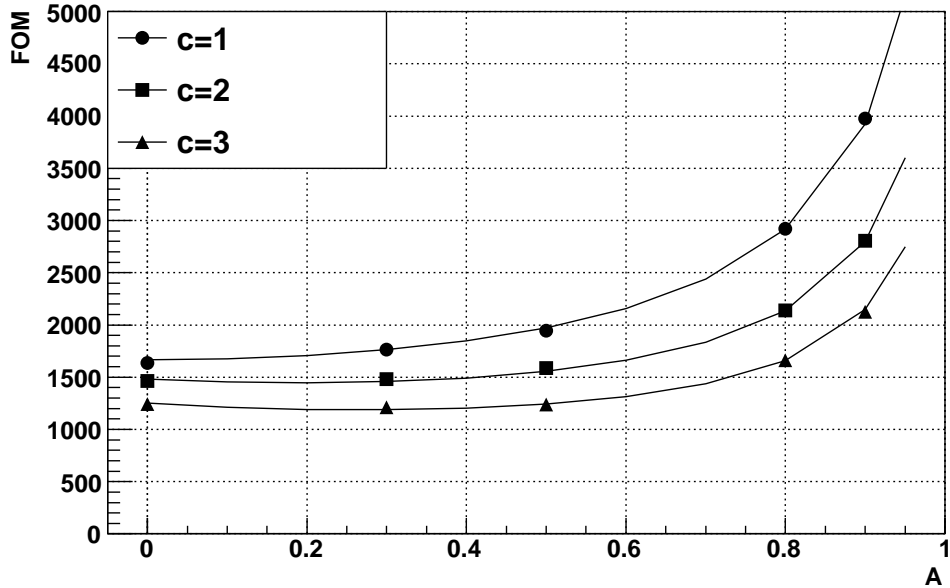}
\caption{Figure of merit for three different values of $c$
as a function of the asymmetry $A$,
assuming the same number of events in the case $A=0$.\label{fom_diff_c}}
\end{figure}

The factor $2/(1+c)$ has been introduced in order to normalize the distributions
to the same number of events for all values of $c$ at $A=0$.
The log likelihood function reads
\begin{eqnarray}
  l &=& \sum_+ \ln \left(\frac{2c}{1+c} \, \alpha_i (1 + \beta_i
  A) \right)- \avrg{N^+}(A)+ \nonumber \\  
                  && \sum_- \ln \left( \frac{2}{1+c} \, \alpha_i (1 - \beta_i A) \right)-  \langle N^-\rangle (A)  \nonumber \\
      &=& \sum_+ \ln \left(1 + \beta_i A \right)  + \sum_- \ln \left(1 - \beta_i A \right)   \, \nonumber  \\
                && +  \sum_{+} \ln \frac{2c}{1+c} \, \alpha_i \, +
                  \sum_- \ln \frac{2}{1+c} \, \alpha_i  \nonumber \\ && 
               - 2 \int \alpha {\rm d} x - 2 A \, \frac{c-1}{1+c} \, \int \alpha
               \beta {\rm d} x    \, .              
\end{eqnarray}
Here the last term cannot be ignored because it contains the parameter $A$
and thus the likelihood method cannot be applied without knowledge of the factor $\int \alpha \beta {\rm d} x$.
The weighting method on the other hand can be applied with a small modification:
\begin{equation}\label{aiw_mod}
  \hat A_{w,c} = \frac{\sum_+ w_i - c \sum_- w_i}{\sum_+ w_i \beta_i +
    c \sum_- w_i \beta_i} \, .
\end{equation}
The expectation value of $\hat A_{w,c}$ equals again $A$.
The figure of merit reads (derivation see s. App.~\ref{app_fomc})
\begin{equation}\label{fom_iw_mod}
   \mbox{FOM}_{\hat A_{w,c}} = %\frac{8c}{(1+c)^2} \, 
%\frac{\left( \int \frac{\alpha \beta^2 {\rm d} x}{1-\beta^2 A^2 }   \right)^2}
%{ \int \frac{\alpha \beta^2}{1-\beta^2 A^2 }  \left(1- \beta A
%  \frac{1-c}{1+c}\right) {\rm d} x} \, .
%N \, \frac{4c}{(1+c)^2} \, \frac{\avrg{\frac{\beta^2}{1-\beta^2
%      A^2}}^2}{ \avrg{\frac{\beta^2}{1-\beta^2 A^2} \left( 1- \beta A
%    \frac{1-c}{1+c}\right)}} \, .
N \, \frac{4c}{(1+c)^2} \, \frac{\avrg{w \beta}^2}{ \avrg{ w^2 (1-\beta^2 A^2) \left( 1- \beta A
    \frac{1-c}{1+c}\right)}} \, .
\end{equation}
The FOM is shown in Fig.~\ref{fom_diff_c} for different values of $c$ for the
improved weighting method.
As in Fig.~\ref{fig_fom} the lines correspond to an analytic calculation
using Eq.~(\ref{fom_iw_mod}), the points are results of simulations.
Note, that for arbitrary $c$ the weight reaching the
highest FOM is 
\[
 w = \frac{\beta}{(1-\beta^2 A^2)\left( 1- \beta A \frac{1-c}{1+c}\right)} \, .
\]
The EML method could be used for $c\ne1$, if one uses an estimate for $\int \alpha \beta {\rm d} x \approx
\sum_+ \beta(x_i) + c \sum_- \beta(x_i)$ from data. 
Simulations showed that the FOM of this modified EML method equals the one of the improved weighting method.

\section{Validity at low number of events}\label{low_stat}
\begin{figure}
\includegraphics[width=\textwidth]{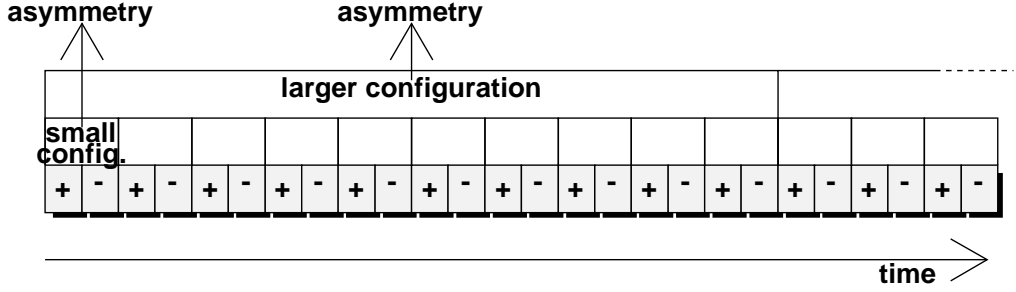}
\caption{Combining data in different configurations. 
The boxes denote the plus and minus data sets.\label{confis}}
\end{figure}

In this section we discuss the validity of the equations derived for the various FOMs
and possible biases of the estimators.
The formulas were derived using usual error propagation and can thus only be approximations
which are the better the higher $N$.
In the simulations presented in Section~\ref{comp}
for every value of the asymmetry  10000 configurations were simulated with on
average 5000 events (corresponding to $\alpha=2500$ in Eq.~(\ref{alpha})).
In each of these configurations the asymmetry was determined using the various estimators.
In this section we discuss effects occurring if the asymmetries are extracted in smaller
configurations, as indicated in Fig.~\ref{confis}.

For lower number of events in one configuration
one reaches a point where due to statistical fluctuations
the estimated asymmetry in a given configuration can be larger than 1 or smaller than $-1$. 
In this case the EML method
is no more applicable since the term $(1 \pm \beta A)$
can get negative.
For $\alpha=250$ this happens in about $1\%$ of the configurations 
for an asymmetry of $A=0.8$.

Dividing the sample further in many smaller configurations 
one reaches a point where one has only 0 or 1 event in one configuration.
This limit can also be obtained by dividing the sample
in many narrow bins of $\beta$ having at most one event in a bin.
In this case the remaining estimators are identical: $\hat A_{cnt}
\equiv \hat A_{w = \beta} \equiv \hat A_{iw} = \pm 1/\beta_i$. 
The sign depends whether the event occurred in the plus or the minus
data set. One finds $\avrg{\hat A_{iw}} = A$ and $\avrg{\hat A_{iw}^2} =
1/\beta_i^2$,
thus the FOM for this event reads 
\[
\mbox{FOM}_i = \frac{1}{\avrg{\hat A_{iw}}^2 - \avrg{\hat A_{iw}^2}} = \frac{\beta_i^2}{1- \beta_i^2 A^2} \, . 
\]
Note that $A$ is not the estimated asymmetry from a single event but rather
taken from a larger event sample.
Combining all the asymmetries determined on single events leads to
  \begin{equation}\label{a_se}
\frac{\sum_i \pm\frac{1}{{\beta}_i} \,\cdot \, \mbox{FOM}_i}{\sum_i \mbox{FOM}_i} = 
  \frac{\sum_+ \frac{1}{\beta_i} \, \frac{\beta_i^2}{1-\beta_i^2 A^2}- \sum_- \frac{1}{\beta_i} \, \frac{\beta_i^2}{1-\beta_i^2 A^2}}
      {\sum_+ \frac{\beta_i^2}{1-\beta_i^2 A^2}+ \sum_- \frac{\beta_i^2}{1-\beta_i^2 A^2}} \, . 
\end{equation}
Assuming $A = A_0$, Eq.~(\ref{a_se}) is nothing but the estimator of the improved
weighting method, $\hat A_{iw}$, defined in Eq.~(\ref{a_iw}).
In other words, for the improved weighting method it makes no differences
whether the data are analyzed in one large configuration or in many small ones.
The improved weighting is also equivalent to
using an infinite number of bins in $\beta$
and evaluating the asymmetries
in every bin and then combining the results.
The advantage of the improved weighting method is that this
binning has not to be performed.

The observations discussed above are confirmed by simulations.
In total $10^9$ configurations with $\alpha = 0.025$ were simulated.
The simulated data were analyzed as follows.
First the asymmetries were calculated in the approximately
5\% of the configurations actually containing at least one event. Then the
weighted average of these asymmetries is calculated. 
The same data were analyzed in a different way
by combining 10 configurations and calculating the asymmetries 
in these larger configurations corresponding to $\alpha = 0.25$.
This procedure was repeated until reaching $10^4$ configurations with $\alpha=2500$.
Fig.~\ref{confis} illustrates the procedure.
These simulations were performed for an asymmetry of 0.8 
generating events in the range $0.01<\beta(x)<0.99$ .

\begin{figure}[h!t]
\includegraphics[width=0.9\textwidth]{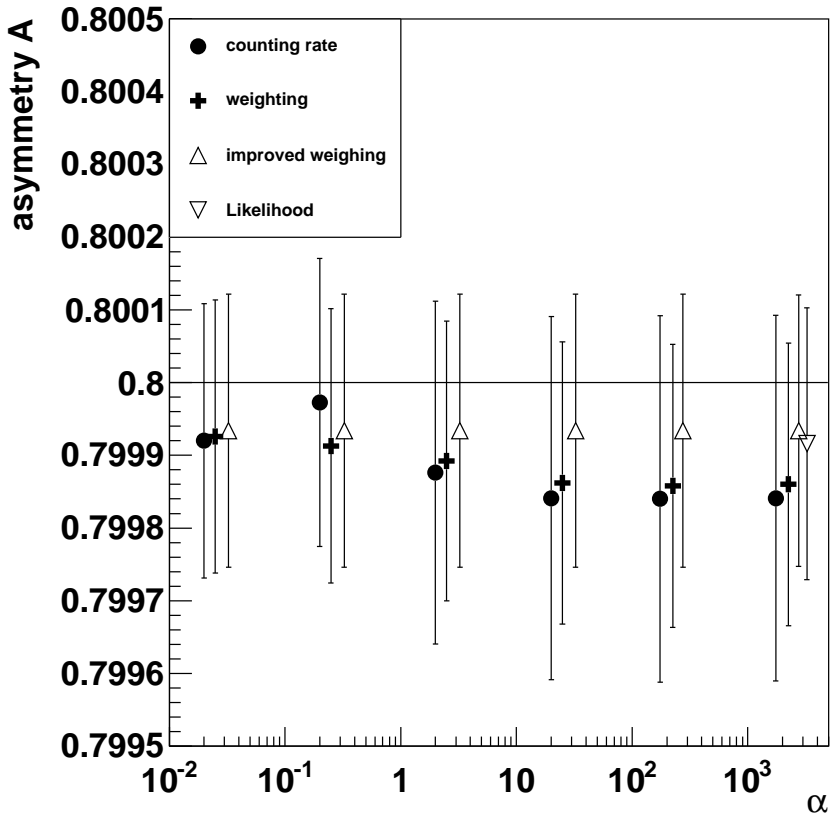}
\caption{Results for the asymmetry of the simulations as a function
of the average number of events in one configuration.
The points are at $\alpha = 0.025, 0.25, 2.5, 25, 250, 2500$, respectively.
For a given value of $\alpha$ they are slightly displaced on the horizontal axis for
better readability. Note, that values at different values of $\alpha$ are correlated
since the same data were used.\label{bias_asy}}
\end{figure}

Fig.~\ref{bias_asy} shows the mean value of the asymmetries and the
statistical error for the various estimators for the different values of
$\alpha$.
As expected the estimator $\hat A_{iw}$ gives the same result independent of $\alpha$.
No bias is observed  within the statistical error which is
of the order of $10^{-4}$.
The asymmetry for the EML method is only shown for $\alpha=2500$ since at lower 
values, as explained above, the EML method is no more applicable.

\begin{figure}[h!t]
\includegraphics[width=0.9\textwidth]{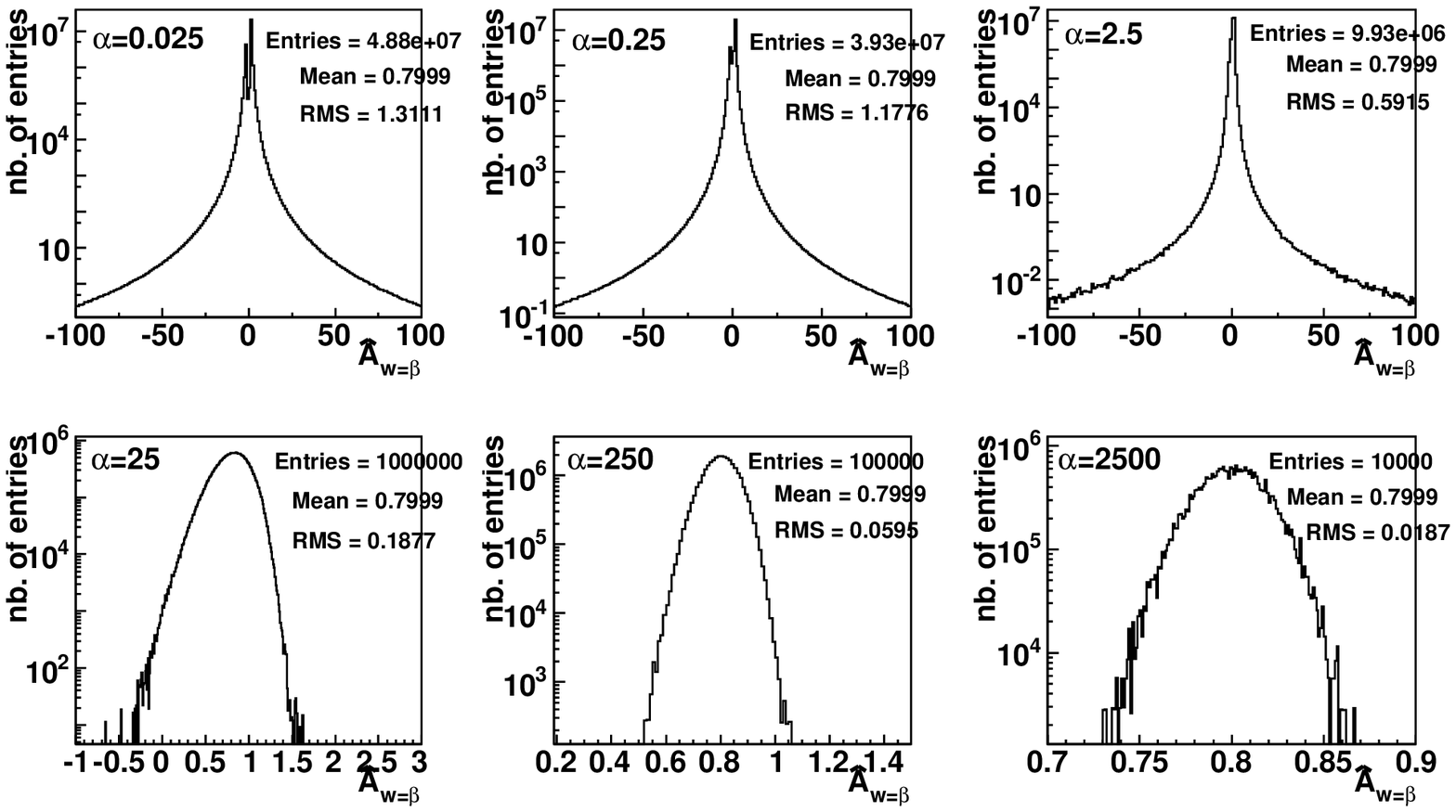}
\caption{Distributions of the estimated asymmetries $\hat A_{iw}$ for
  different values of $\alpha$.
\label{bias_asy1}}
\end{figure}

Fig.~\ref{bias_asy1} shows the distribution of the asymmetry $\hat A_{iw}$
for different values of $\alpha$.
%The mean and the RMS/$\sqrt{\mbox{Entries}}$ correspond to the values shown in
%Fig.~\ref{bias_asy} for $\hat A_{iw}$.
The entries in the histograms in Fig.~\ref{bias_asy1} are weighted by their 
corresponding FOM. In the case $\alpha=2500$ this would not be necessary
because all entries have essentially the
same FOM for a given method since the relative variation of the number of events
$N$ and the averages like $\sum_\pm \beta^2_i/N$ entering the FOM vary only very little from
configuration to configuration.
At lower values of $\alpha$, however, this is no more the case.
Taking again the extreme case where the asymmetries is calculated from single
events, the FOM depends on the value of $\beta$ for this event.
This explains why the number of entries is smaller than 1 in some bins of the histograms. 
At $\alpha=0.025$ the number of configurations is $10^9$.
The corresponding histogram has only approximately $4.9 \cdot
10^7$ entries reflecting the fact that in most of the configuration
there is no event.

\begin{figure}[h!t]
\includegraphics[width=0.9\textwidth]{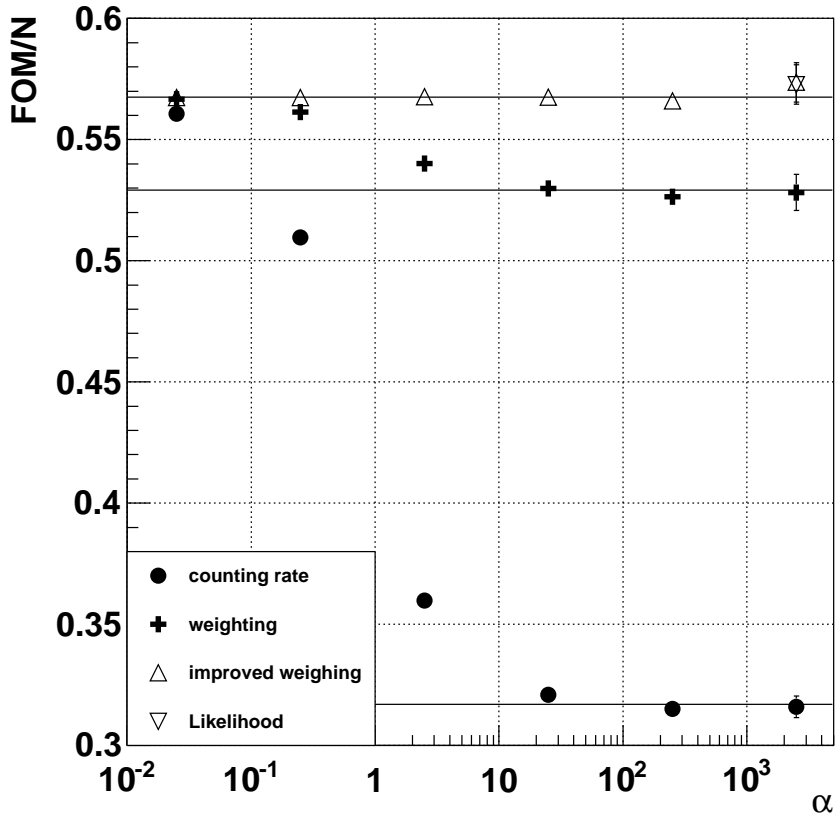}
\caption{Figure of merit per event for different methods as a function of $\alpha$
for an asymmetry $A=0.8$. The lines show the expectation calculated from
the expressions given in Tab.~\ref{tab_fom}.
\label{ratio_fom}}
\end{figure}

Finally, Fig.~\ref{ratio_fom}
shows the FOM/$N$ calculated from the RMS of the
asymmetry distributions presented in Fig.~\ref{bias_asy1} for different values
of $\alpha$.
The three lines correspond to $\mbox{FOM}_{\hat A_{iw}}/N$,
$\mbox{FOM}_{\hat A_{w=\beta}}/N$ and $\mbox{FOM}_{\hat A_{cnt}}/N$
calculated using the expressions given in Tab.~\ref{tab_fom}.
For $\alpha \ge 25$ there is good agreement with
the FOM derived in Section~\ref{diff_methods} since the points coincide
with the corresponding lines.
At lower values of $\alpha$ the FOM of the weighting and the counting rate method start to increase 
and finally reach as expected $\mbox{FOM}_{\hat A_{iw}}$ at $\alpha = 0.025$,
where the three estimators are practically identical.

\section{Summary \& Conclusions}\label{sum}
We presented several estimators to extract an asymmetry parameter $A$
in a number density function.
These estimators were based on counting rates, event weighting
and the unbinned extended maximum likelihood method.
A weighting procedure was derived that reaches the same figure of merit
as the unbinned maximum likelihood method, known to reach the minimal variance bound.
This weighting estimator is given as an analytic expression,
whereas in the EML method the maximization of the likelihood function
has to be done numerically.
Moreover this estimator can be used (with a small modification)
in cases where the EML method cannot be applied 
because of an incomplete knowledge of event distribution function.

Acknowledgments: I am grateful to Jean-Marc Le Goff
for numerous discussions on the subject, verifying the calculations and for carefully reading the manuscript.
\clearpage

%-------------------------------------------------------------------------------
\appendix
\section{Figure of merit of $\hat A_w$}\label{app_covwb}
To calculate the FOM of $\hat A_w$ defined in Eq.~(\ref{aw}),
one needs $\sigma^2(\sum w_i)$, $\sigma^2(\sum w_i \beta_i)$ and $\mbox{cov}(\sum w_i, \sum w_i \beta_i)$.
For two arbitrary quantities $f$ and $g$ the covariance between $\sum f_i$
and $\sum g_i$ is
\begin{eqnarray}
   \lefteqn{\mbox{cov}(\sum_i  f_i, \sum_j  g_j )}\\  
%   &=& \avrg{ \sum_i g_i \sum_j f_j} - \avrg{ \sum_i f_i} \avrg{\sum_j g_j}   \\
                                  &=& \avrg{ \sum_{i=j} f_i g_i + \sum_{i\ne j} f_i g_j} 
                                       - \avrg{ \sum_i f_i} \avrg{\sum_j g_j}  \\
                                   &=& \avrg{N}  \avrg{f g} + \avrg{N(N-1)}  \avrg{f}  \avrg{g} - \avrg{N}^2 \avrg{f} \avrg{g}  \\
                                   &=& \avrg{N}  \avrg{f g} + \left(\avrg{N^2} - \avrg{N} - \avrg{N}^2\right)  \avrg{f}  \avrg{g}   \, .
\end{eqnarray}
If the number of events $N$ is Poisson distributed, i.e.
$\avrg{N^2} - \avrg{N} - \avrg{N}^2 =0$,
one finds 
\begin{equation}
\mbox{cov}(\sum_i f_i, \sum_j g_j) = \avrg{N}  \avrg{f g}  \approx \sum_i f_i g_i \, .
\end{equation}

Setting $f=g=w$, $f=g=w\beta$ and $f=w, g=w \beta$ results in 
\begin{eqnarray}
\sigma^2(\sum w_i) &=& \avrg{N} \, \avrg{w^2} \label{sig_w} \, ,\\
\sigma^2(\sum w_i \beta_i) &=& \avrg{N} \, \avrg{(w \beta)^2} \label{sig_wb}
\, ,\\
\mbox{cov}(\sum w_i, \sum w_i \beta_i) &=& \avrg{N} \, \avrg{w^2 \beta
} \label{cov_wwb} \, .
\end{eqnarray} 
Simple error propagation in Eq.~(\ref{aw}) finally leads to Eq.~(\ref{fom_aw})
for the figure of merit.

\section{Optimal weight}\label{vc}
Denoting the weight factor which maximizes the FOM by $w_0$,
we consider small deviation from this optimum by 
\begin{equation}\label{opt_w}
  w(x) = w_0(x) + \epsilon \,\eta(x)
\end{equation}
where $\eta(x)$ is arbitrary and $\epsilon \ll $1.

Inserting Eq.~(\ref{opt_w}) in Eq.~(\ref{fom_aw}) keeping terms
of 1st order in $\epsilon$ one finds
\begin{equation}
  \mbox{FOM} = \frac{\left( \avrg{w_0 \beta} + \epsilon \avrg{\eta \beta}
    \right)^2}{\avrg{\left(w_0^2 + 2\epsilon w_0 \eta \right)(1-\beta^2 A^2)}} \, .
\end{equation}
The condition $\partial \mbox{FOM}/\partial{\epsilon} = 0$
gives 
\[
  w_0 = \frac{\beta}{1-\beta^2 A^2} \, .
\]

\section{FOM for the case $c\ne 1$}\label{app_fomc}

The error for the estimator defined in Eq.~(\ref{aiw_mod}) is obtained by simple
error propagation taking into account the correlations between $\sum w \beta$
and $\sum w$.
\[
 \left(\mbox{FOM}_{\hat A_{w,c}} \right) ^{-1} = \vec v^T \, C \,  \vec v \, 
\]
with
\begin{eqnarray}
 \vec v^T &=& \left(\frac{\partial \hat A_{w,c}}{\partial (\sum_+ w_i)},
                  \frac{\partial \hat A_{w,c}}{\partial (\sum_- w_i)},
                  \frac{\partial \hat A_{w,c}}{\partial (\sum_+ w_i \beta_i)},
                  \frac{\partial \hat A_{w,c}}{\partial (\sum_- w_i \beta_i)},
  \right) \nonumber \\
    &=& \frac{1}{\sum_{+}  w_i \beta_i + c \sum_{-}  w_i \beta_i} \,
     \left(1,-c,-A,-cA\right) 
\end{eqnarray}
and
\begin{eqnarray}
C= \left(
\begin{array}{cccc}
\sum_+ w_i^2 & 0 & \sum_+ w_i^2 \beta_i & 0 \\
0  & \sum_- w_i^2 & 0 & \sum_- w_i^2 \beta_i \\
\sum_+ w_i^2 \beta_i & 0 & \sum_+ (w_i \beta_i)^2 & 0 \\
0 & \sum_- w_i^2 \beta_i & 0 & \sum_- (w_i \beta_i)^2 \\
\end{array}
\right) \, .
\end{eqnarray}
This leads to Eq.~(\ref{fom_iw_mod}).

%From Eq.~(\ref{aiw_mod}) the figure of merit is obtained (neglecting terms proportional to $\Delta$)
%\begin{equation}
%  \mbox{FOM}_{\hat A_{iw}} = \frac{\left( \sum_+ (\beta^+_i)^2 + c \sum_- (\beta^-_i)^2\right)^2}
%           {\sum_+ (\beta^+_i)^2 + c^2 \sum_- (\beta^-_i)^2} \, .
%\end{equation}
%The expectation values $\avrg{\sum_+ (\beta^+_i)^2}$ and $\avrg{\sum_- (\beta^-_i)^2}$ are
%\begin{eqnarray}
%\avrg{\sum_+ (\beta_i^+)^2} &\stackrel{\Delta=0}{=}&   \frac{2c}{1+c} \int \alpha^+ (\beta^+)^2 {\rm d} x
%\stackrel{\alpha^+ \beta^+ = \alpha \beta}{=} \frac{2c}{1+c} \int \alpha \beta \beta^+ {\rm d} x
%\quad \mbox{and} \nonumber \\ 
%\avrg{\sum_- (\beta_i^-)^2} &\stackrel{\Delta=0}{=}& \frac{2}{1+c} \int \alpha \beta \beta^- {\rm d} x \, . \nonumber
%\end{eqnarray}%

%Thus
%\begin{equation}
%  \mbox{FOM}_{iw} = \frac{2c}{1+c} \, \frac{ \left( \int \alpha \beta (\beta^+ + \beta^-)\right)^2}
%                      {\left( \int \alpha \beta (\beta^+ + c\beta^-)\right) }
%                      \, .
%\end{equation}
%Using
%\begin{eqnarray}
%  \beta^+ + \beta^- &=& \frac{2 \beta}{1-\beta^2 A^2} \quad 
%\mbox{and} \nonumber \\    
%  \beta^+ + c \beta^- &=& \frac{(1+c)\beta}{1-\beta^2 A} \, \left( 1 + \beta A
%  \frac{c-1}{c+1}\right) \nonumber
% \end{eqnarray}
%leads to Eq.~(\ref{fom_iw_mod}).

%%%%%%%%%%%%%%%%%%%%%%%%%%%%%%%%%%%%%%%%%%%%%%%%%%%%%%%%%%%%%%%%%%%%%%%%%

\end{document}